\documentclass[landscape]{aa}
\usepackage{graphicx,psfrag,amsmath,lscape}
\begin{document}

\title{The Top Ten solar analogs in the ELODIE library
\thanks{Based on observations  made at the Observatoire
        de  Haute-Provence (France)}}
\titlerunning{Solar analogs}
\author{Soubiran C.\inst{1} \& Triaud A.\inst{1,2}}
\offprints{C. Soubiran}
\mail{Caroline.Soubiran@obs.u-bordeaux1.fr }
\institute{Observatoire Aquitain des Sciences de l'Univers,  UMR 5804, 2  
rue de  l'Observatoire,  33270
              Floirac, France \and
	      University of St-Andrews, School of Physics and Astronomy, North Haugh,
	       St Andrews, Fife KY16 9SS, Scotland}
\date{Received \today / Accepted }
\abstract{Several solar analogs have been identified in the library of high 
resolution stellar spectra taken with the echelle
spectrograph ELODIE. A purely differential method has been used, based on the $\chi^2$ comparison of a
large number of G dwarf spectra to 8 spectra of the Sun, taken on the Moon and Ceres. HD 146233 keeps its
status of closest ever solar twin (Porto de Mello \& da Silva, \cite{PMDS97}). 
Some other spectroscopic analogs 
have never been studied before, while the two planet-host stars HD095128 and HD186427 are also part of the 
selection. The fundamental parameters found in the literature for these stars
show a surprising dispersion, partly due to the uncertainties which affect them. We discuss
the advantages and drawbacks of photometric and spectroscopic methods to search for solar analogs and 
conclude that they have to be used jointly to find real solar twins.
}
\maketitle

\section{Introduction}
The Sun is the best-known star : its fundamental parameters (radius, mass, age, luminosity,
effective temperature, chemical composition) are known with a good accuracy, as well as its
internal structure, activity, velocity field and magnetic field. Consequently the Sun is used as
the fundamental standard in many astronomical calibrations. One of the motivations to
identify stars that replicate the solar astrophysical properties is the necessity to
have other reference stars, observable during the night under the same conditions as any other 
target. The pioneers of the subject 
(Hardorp \cite{H78}, Cayrel de Strobel et al \cite{cay81}) 
were also involved in resolving the problem of the photometric indexes of the Sun, inherent to the 
impossibility to observe it as a point-like source. In the last decade the motivation of
finding such stars has been increased by an exciting challenge : the search for planetary systems
that could harbour life. Solar analogs are straightforward targets for this hunt. 

The first searches of solar analogs were performed by photometric and spectrophotometric techniques.
Hardorp (\cite{H78}) compared UV spectral energy distributions of nearly 80 G dwarfs to that
of the Sun and found 4 stars that had a UV spectrum indistinguishable from solar : HD028099 (Hy VB 64),
HD044594, HD186427 (16 Cyg B), HD191854. Neckel (\cite{neck86}) established a list of bright stars 
with UBV-colours close to those of the Sun and confirmed the photometric resemblance of Hy VB 64 and 
16 Cyg B to the Sun. With the advance of 
techniques in high resolution spectroscopy and solid state detectors, and with the progress
in modelling stellar atmospheres, measurements of (T$_{\rm eff}$, log\,g, [Fe/H]) became
of higher precision allowing the search for solar analogs by comparing their atmospheric parameters
to those of the Sun. G. Cayrel de Strobel made a huge contribution to the subject with the detailed
analysis of many candidates (Cayrel de Strobel et al \cite{cay81}, Cayrel de Strobel \& Bentolila \cite{cay89},
Friel et al \cite{friel93}) and a review of the status of the art (Cayrel de Strobel \cite{cay96}). She
also introduced the concepts of solar twin, solar analog and solar-like star.
 Porto de Mello \& da Silva 
(\cite{PMDS97}) presented the star HD146233 (18 Sco) with physical properties extremely close to 
those of the Sun, as the "closest ever solar twin". 

A workshop on Solar Analogs was held  in 1997 at the Lowell Observatory to provide a 
solid basis to the hunt of solar analogs. After many discussions on the performances of different methods,
a list of the best candidates was established, in which 4 stars received the agreement of 
almost all participants : HD217014 (51 Peg), HD146233 (18 Sco), HD186408 (16 Cyg A), HD186427 (16 Cyg B).

In this paper, we take advantage of a large and homogeneous dataset of high resolution echelle 
spectra which are compared directly to solar
spectra, independently of any model or photometric measurements. The eye is replaced by a 
more reliable criterion, approximatively the reduced $\chi^2$, computed over $\sim$ 32000 resolution
elements. This purely 
differential method allowed us to identify several stars whose optical spectrum looks closely like 
the Sun's, the best one being HD146233. 

We describe in Sect. \ref{s:ELO} our observational material and differential method, and we give the list of
our Top Ten solar analogs. We have searched the literature for their colour indexes and atmospheric
parameters and calculated absolute magnitudes from Hipparcos parallaxes. We discuss the 
uncertainties which affect these data and compare them to that of the Sun (Sect. \ref{s:param}). In
Sect. \ref{s:Li}, we examine qualitatively their Li content and give information on their activity and
age. In Sect. \ref{s:other} we discuss several 
stars, having similar colours and absolute magnitude or similar atmospheric parameters as the Sun but
slightly different spectra.

\section{ELODIE spectra and the TGMET code}
\label{s:ELO}
All the spectra used in this paper were extracted from the library of stellar spectra collected  with the 
echelle spectrograph ELODIE at the Observatoire de Haute-Provence by Soubiran et al (\cite{SKC98}) and 
Prugniel \& Soubiran (\cite{PS01}). The 
performances of the instrument mounted on the 193\,cm telescope, are described in Baranne et al (\cite{bar96}).
ELODIE is a very stable instrument, built to monitor radial velocity oscillations of stars with exoplanets, 
at a resolving power of 42\,000 in the wavelength range 3850--6800 {\AA}. The stability of the system 
makes it possible to compare spectra taken at different epochs. Spectrum extraction, wavelength calibration and radial 
velocity measurement by cross-correlation have been performed at the telescope with the on-line  data 
reduction software. 

The current version of the Elodie library includes 1962 spectra available in the Hypercat 
database\footnote{www-obs.univ-lyon1.fr/hypercat/11/spectrophotometry.html}, most of the spectra having
a signal to noise ration (S/N) at 550\,nm greater than 
100. We have selected  
208 spectra of G dwarfs with the following criteria : $0.55 < B-V < 0.75$ ($(B-V)_{\odot}\simeq 0.65$) and 
$4 < M_{\rm V} < 5.6$ ($M_{\rm V \odot }\simeq 4.82$). Absolute magnitudes have been computed from Hipparcos 
parallaxes, the selected stars having relative errors smaller than 10\%. The list includes 8 spectra of the Sun 
(Table \ref{t:solar_spectra}). 

	\begin{table}[hbtp]
	\caption{
	\label{t:solar_spectra}
	List of the solar spectra used in this study.}
	\begin{center}
	\begin{tabular}{c c c c c}
	\hline 
	\hline  
	Hypercat  &  date of     & object   &         FWHM          & S/N at  \\ 
	number    &  observation &          &  $\rm{km.s}^{-1}$     & 550\,nm \\
	\hline
	00903 & 14/01/1998 & Moon & 11.061& 381.4 \\
	00904 & 22/12/1999 & Moon & 11.050&268.5\\
	00905 & 22/12/1999 & Moon & 11.050&139.6\\
	00906 & 22/12/1999 & Moon & 11.061 &156.5\\
	00907 & 27/03/2000 & Ceres &11.017& 117.5\\
	00908 & 24/01/2000 & Moon & 11.057&200.0\\
	00909 & 24/01/2000 & Moon &11.054 &224.9\\
	01964 & 22/08/2000 & Moon & 12.126 &404.3\\
	\hline
	\end{tabular}
	\end{center}
	\end{table}

The stellar spectra were compared to solar ones with the TGMET code developed by Katz et al (\cite{KSC98}).
TGMET is a minimum distance method which measures 
similarities between spectra in  a quantitative way. The TGMET comparison between 2 spectra includes 
the following steps :

	\begin{itemize}
	\item{}straightening of each order 
	\item{}removal of bad pixels, cosmic hits and telluric lines 
	\item{}wavelength adjustment
	\item{}mean flux adjustment by weighted least squares, order by order
	\end{itemize}

The wavelength adjustment shifts the comparison spectrum to the radial velocity of the solar spectrum and
resamples it to the same wavelenghts. It
implies an interpolation between wavelengths which is performed with the quadratic Bessel formula.
The flux fitting of the comparison spectrum to the solar spectrum assumes that the 2 spectra differ by a 
simple factor (the 2 stars having roughly the same temperature it is not necessary to introduce a slope).  
Once the two spectra have been put on a common scale, a distance between them 
can then be computed. As explained in Katz et al (\cite{KSC98}), instead of adopting the real reduced 
$\chi^2$ of the fit as the distance between 2 spectra, which would imply taking into account the noise 
on each pixel, the response curve was chosen as the weighting function. This distance was adopted after
many tests and was proven to produce the most satisfactory results, especially at high S/N. 
Its advantage 
is that it gives a similar weight to the continuum and to the wings and bottom of absorption lines, 
contrary to a weighting function based on the photon noise.

Katz et al's algorithm includes a convolution step which is not included in the present work. 
A convolution should be performed in order to put the 2 compared spectra at exactly the same 
resolution. A difference in resolution between two spectra is the result either of a variation of the 
instrumental
resolution between the two exposures or  of intrinsic physical properties of the observed stars 
like
rotation, macroturbulence or binarity which enlarge their spectral profile. But as we are looking for
solar twins, these intrinsic properties are important in the criterion of similarity and should not be
erased. Moreover ELODIE is a very stable
instrument and its resolution does not vary significantly with time. It can be seen in Table 
\ref{t:solar_spectra} that all of our solar spectra have a resolution of 11 $\rm{km.s}^{-1}$ (FWHM), 
except the spectrum 01964 which is sligthly degraded (12 $\rm{km.s}^{-1}$). These considerations led 
us to supress the convolution step in the TGMET algorithm.

 In practice we have limited
the comparison to the wavelength range 4400--6800 {\AA} (orders 21 to 67) and eliminated the under-illuminated 
edges of the orders. Finally distances between spectra have been computed over nearly 32000 wavelengths. 
Table \ref{t:TGMET_Output} gives an example of the TGMET output, for solar spectra 00903 
(S/N=381.4) and 00907 (S/N=117.5). The output in the two cases is consistent, with however some
differences  :  HD088072 is within the 20 closest neigbours of 00903 but not of 00907, 
the opposite is the case for HD071148 and HD042618. These
differences, probably related to observing conditions, are smoothed when combining 
the TGMET results obtained for the 8 solar spectra, the combination being performed by
averaging the distances, order by order, giving a different weight to several orders (see below).
The score obtained by each solar analog was our criterion of closeness to the Sun, leading the 
final Top Ten list : HD146233,  HD168009, HD010307, HD089269, HD047309, 
HD095128, HD042618, HD071148, HD186427, HD010145.

\begin{figure*}[hbtp]                                 
	\begin{center}
	\includegraphics[width=8cm]{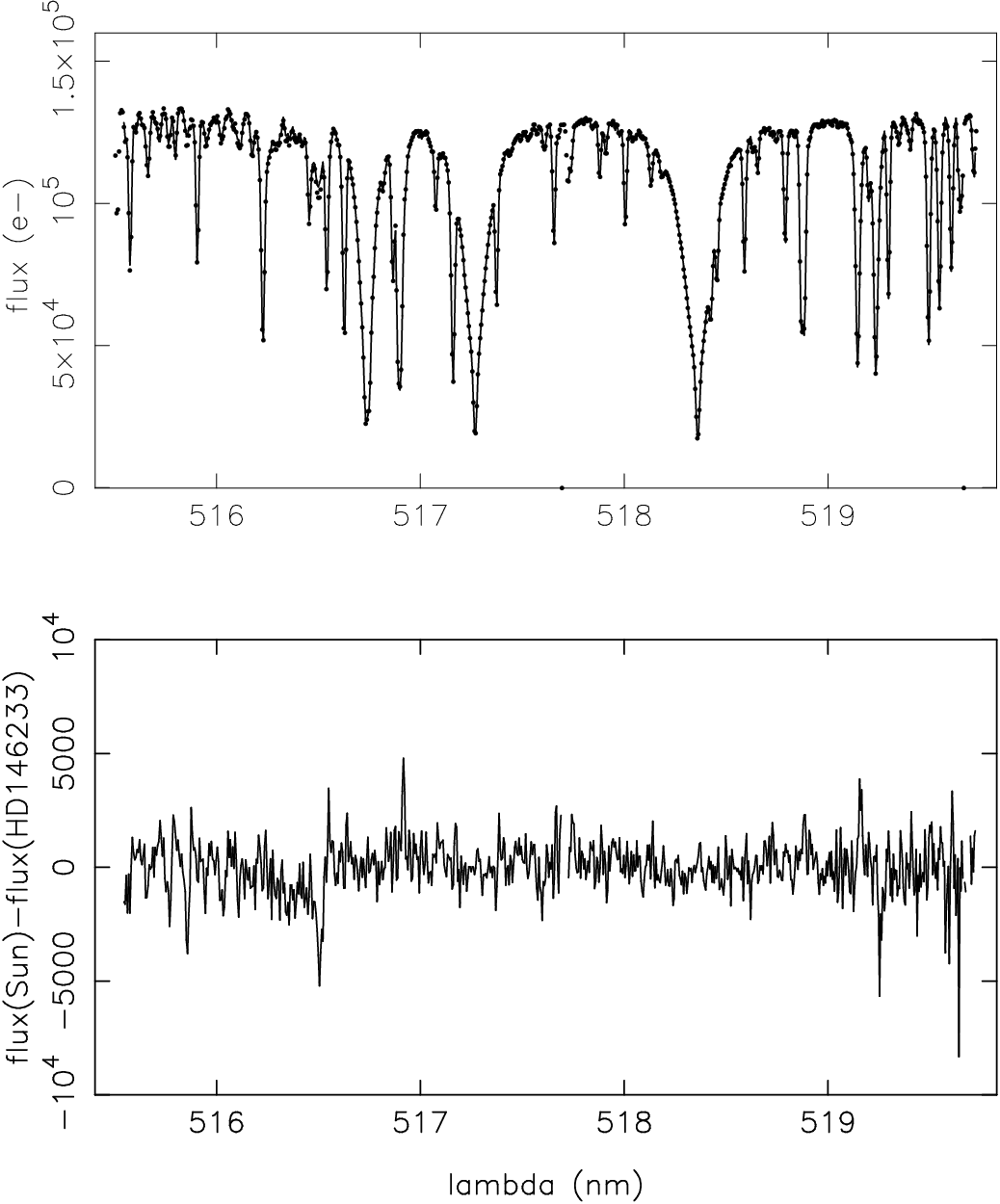}     
	\includegraphics[width=8cm]{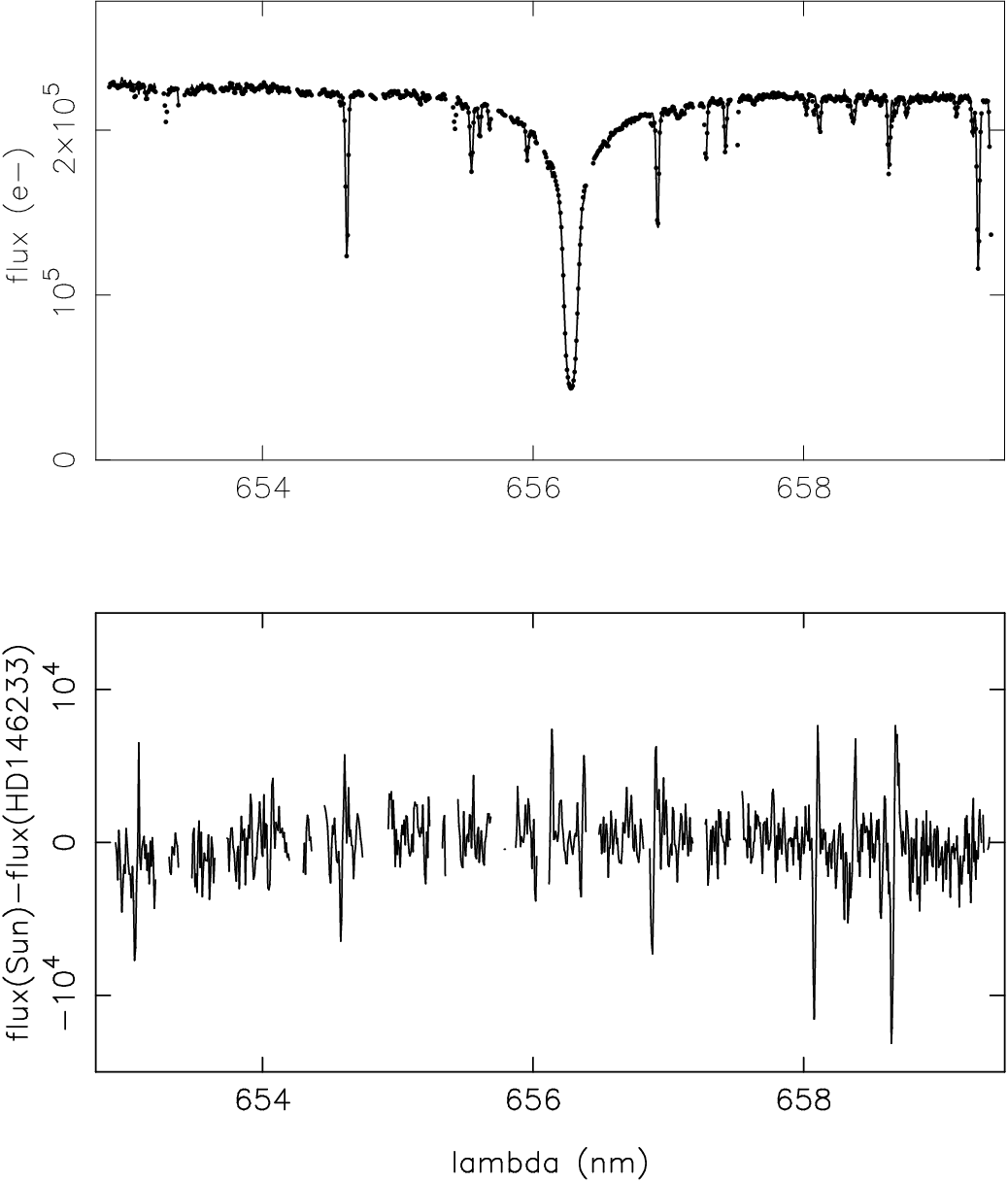}     
	\caption{Comparison of one of the solar spectra (dots) with HD146233 (continuous line) 
	in the spectral region of the MgI triplet and $H_{\alpha}$ line.}  
	\label{f:MgHacomp} 
	\end{center} 
	\end{figure*}

Fig.
\ref{f:MgHacomp} shows the fit of HD146233 to spectrum 00903, for order 39 including 
the MgI triplet and order 64 including the $H_{\alpha}$ line. For order 39, the fit has been 
performed on 773 points; the mean difference between the solar flux and the fitted fluxes of HD146233 
is 3 electrons with a standard deviation of 1157 electrons, corresponding to 1\% of the mean flux (108772 
electrons). For order 64, the difference is also 1\% (743 points, mean difference of -\,57 electrons, 
standard deviation of 2166 electrons, mean flux of 210319 electrons).

	\begin{table*}[hbtp]
	\caption{
	\label{t:TGMET_Output} 
 	The 20 closest spectra of the solar spectra 00903 and 00907, deduced by the TGMET code.}
	\begin{center}
	\begin{tabular}{c c r c | c c r c}
	\hline
	\hline
 	n$^{\rm o}$  & star & S/N & distance & n$^{\rm o}$  & star & S/N & distance \\
	\multicolumn{4}{c |}{00903} & \multicolumn{4}{c}{00907} \\
	\hline
	 01964 & Sun & 404.3 &2.220 & 00903 & Sun & 381.4 & 1.158\\
	 00909 & Sun & 224.9 &2.332 &00909 & Sun & 224.9 &1.175\\
	 00908 & Sun & 200.0 &2.410 &01964 & Sun & 404.3 &1.183\\
	 00490 & HD146233& 236.8 &2.789 &00908 & Sun & 200.0 &1.188\\
	 00906 & Sun& 156.5 &2.910 &00490 & HD146233& 236.8 &1.243\\
	 00905 & Sun& 139.6 &3.110 &00906 & Sun& 156.5 &1.250\\
	 01633 & HD168009& 204.9 &3.305 &00905 & Sun& 139.6 &1.275\\ 
	 00904 & Sun& 268.5 &3.392 &01633 & HD168009& 204.9 &1.329\\
	 00907 & Sun & 117.5 &3.474 &00904 & Sun& 268.5 &1.382\\
	 01187 & HD047309& 119.7 &3.841 &00039 & HD010307 & 198.4 &1.457\\
	 01188 & HD047309& 108.8 &3.864 &01634 & HD168009 & 134.6 &1.458\\
	 01634 & HD168009 & 134.6 &3.875 &01187 & HD047309& 119.7 &1.471\\
	 00039 & HD010307 & 198.4 &3.937 &00346 & HD071148& 117.0 &1.472\\
	 00895 & HD089269 & 225.9 &3.978 &00895 & HD089269 & 225.9 &1.478\\
 	 00258 & HD047309 & 100.9 &4.176 &01188 & HD047309& 108.8 &1.493\\
	 00387 & HD088072 &  86.5 &4.236 &00400 & HD095128 & 181.9 &1.555\\
         00699 & HD186427 & 139.9 &4.283 &00258 & HD047309 & 100.9 &1.557\\
         00400 & HD095128 & 181.9 &4.336 &00981 & HD010307 & 162.9 &1.560\\
         00981 & HD010307 & 162.9 &4.342 &01125 & HD042618 & 132.5 &1.591\\
         00038 & HD010145 & 153.2 &4.346 &00699 & HD186427 & 139.9 &1.594\\
	\hline
	\end{tabular}
	\end{center}
	\end{table*}

\begin{figure*}[hbtp]
	\begin{center}
	\includegraphics[width=16cm,height=10cm]{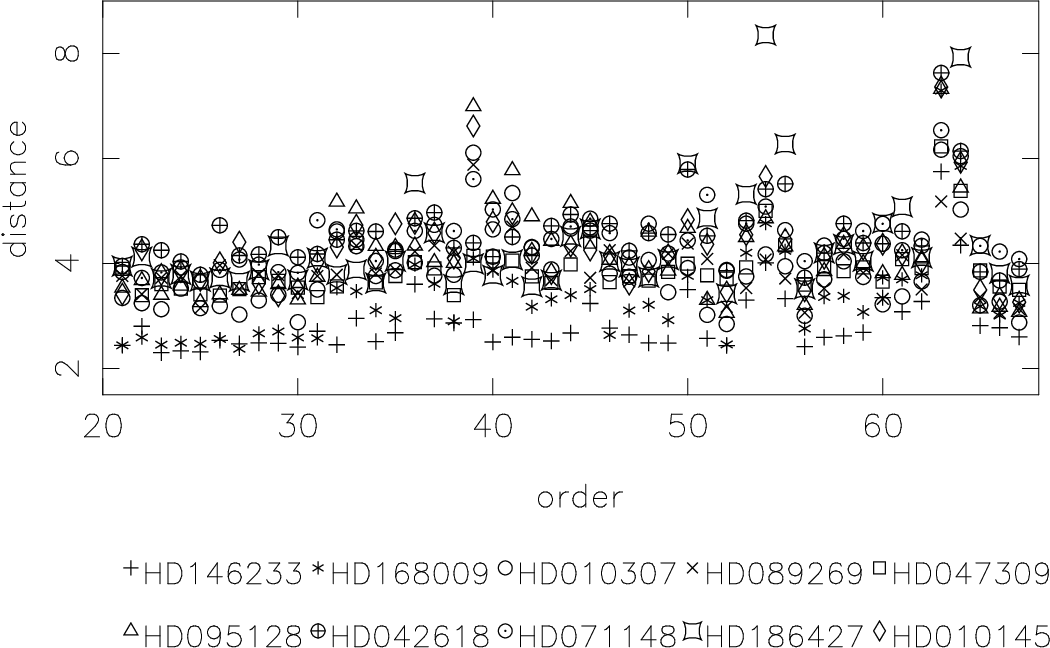}     
	\caption{Distance of the Top Ten solar analogs to solar spectrum 00903, order by order. }  
	\label{f:ki2}  
	\end{center} 
	\end{figure*}

Fig. \ref{f:ki2} shows for the Top Ten solar analogs their distance to the solar spectrum 00903, 
order by order.
It is very clear from  Fig. \ref{f:ki2} that HD146233 is very similar to the Sun and that HD168009 
is not very far behind. The closeness of these 2 stars is confirmed for the 7 other solar spectra. The largest 
discrepancies occur for order 63 (648\,-\,652.5\,nm). The examination of this order
indicates that it is polluted by telluric lines which were not completely removed. Telluric lines
unfortunatly affect also order 64 which is our best indicator of temperature thanks to the $H_{\alpha}$ 
line. The dispersion obtained on order 64 is much higher than that on order 31 which 
includes the  $H_{\beta}$ line. However, the $H_{\beta}$ line being at the edge of the order it has a 
lower weight in the fit because of under-illumination. A large dispersion on order 39, which includes 
strong features due to the MgI triplet, is also seen for 
each solar spectrum. This region is known to be very sensitive to the 3 atmospheric parameters 
(T$_{\rm eff}$, log\,g, [Fe/H]) and consequently powerful for discriminating solar twins.  These considerations 
led us to adopt a higher weight 
of 3 on order 39, a half weight on order 64 and a null weight on order 63 when combining the information on 
all the orders.

\section{Atmospheric parameters and photometry}
\label{s:param}
We report in this section colour indexes available
in the literature for our Top Ten solar analogs and visual absolute magnitudes deduced 
from their Hipparcos
parallaxes (Table \ref{t:general_table}).  We also review recent
determinations of their atmospheric parameters. These data are compared to those
of the Sun, and  uncertainties which affect them are discussed. Finally we report their occurrence 
in previous studies of solar analogs. 

The $B-V$ and $U-B$ colours come from the General Catalogue of Photometric Data 
(Mermilliod et al \cite{mer97}), except for HD047309 for which we have taken the $B-V$ colour 
from Tycho2 (H{\o}g et al \cite{HOG00}) transformed to the Johnson system. The b-y colours 
are extracted from
the catalogue by Hauck \& Mermilliod (\cite{hau98}).

 The photometry of the Sun is a matter of debate. Neckel 
(\cite{neck86}) has determined $(B-V)_{\odot}=0.650 \pm 0.005, (U-B)_{\odot}=0.195 \pm 0.005, 
M_{\rm V \odot}=4.82 \pm 0.025$, values which are adopted as basic solar data in Allen's Astrophysical
Quantities (2000). Cayrel de Strobel (\cite{cay96}) gives a compilation of solar $(B-V)$ colours
measured by different techniques and determines from the relations colour vs Teff :
$(B-V)_{\odot}=0.642 \pm 0.004, (b-y)_{\odot}=0.404 \pm 0.005$. Porto de Mello \& da Silva (\cite{PMDS97})
obtain with a similar method : $(B-V)_{\odot}=0.648 \pm 0.006, (U-B)_{\odot}=0.178 \pm 0.013$.
We list in Table \ref{t:general_table} a reasonable range of values for the Sun's colours and
absolute magnitudes.

\subsection{HD146233}

HD146233 (18 Sco) was adopted at the Solar Analogs workshop at Lowell Observatory as
one of the best solar twins. Our study confirms with independent data and methods the result 
of Porto de Mello \& da Silva (\cite{PMDS97})
quoting HD146233 as THE closest ever solar twin. In the optical range, its spectrum is indistinguishable
from that of the Sun (Fig. \ref{f:MgHacomp}). Before that, Hardorp (\cite{H78}) using UV spectrophotometry mentioned 
this star
as a solar analog but with the comment "spectrum similar to solar, some absorption features weaker". However, 
this
study was based on a single low-resolution spectrum. This discrepancy is discussed by Porto de Mello 
\& da Silva (\cite{PMDS97}) and by Cayrel de Strobel (\cite{cay96}).

Only 2 determinations of atmospheric parameters are available for this star, in very good
agreement, giving solar values within the error bars. HD146233 seems to be more luminous than
the Sun by 0.05 mag. Its parallax is very accurate ($\pi=71.30 \pm 0.89$ mas), but one may legitimely 
wonder if its photometry is sufficiently accurate to consider this excess of luminosity real.
 As a matter of fact HD146233 
is part of the Catalogue of suspected variables by Kukarkin et al (\cite{kuk81}) who found
a possible amplitude of 0.11 mag.  V magnitudes, measured by several authors between 1957 and
1978 and available in the GCPD (Mermilliod et al \cite{mer97}), range effectively between V=5.48 
and V=5.56. The average $V=5.504 \pm 0.015$ was used to compute an absolute magnitude of $M_{\rm V}=4.77$.
Thus if HD146233 is slightly variable, a higher luminosity than the Sun
cannot be clearly established. But more recently HD146233 was identified by 
Adelman (\cite{ade01}) to be part of the 681 most photometrically stable stars during the 5 years of
the Hipparcos mission, with an amplitude of 0.01 mag. Its median magnitude in the Hipparcos system is 
$H_p=5.6265 \pm 0.0005$. According to the photometric transformation calibrated by Harmanec 
(\cite{har98}), the corresponding apparent visual magnitude is $V=5.493 \pm 0.003$ which confirms 
its higher luminosity.

\subsection{HD168009}

HD168009 has been quite well studied but has never been mentioned as a good solar analog, 
despite its
being part of the list of "bright stars with UBV-colours close to those of the Sun" established by 
Neckel (\cite{neck86}). According
to its spectroscopic gravity and absolute magnitude, HD168009 is more luminous than the Sun. Its
absolute magnitude $M_{\rm V}=4.52$ is quite reliable and relies on a parallax of $\pi=44.08 \pm 0.51$ mas. 
Several estimations of its apparent visual magnitude are in good agreement : V=6.295
according to the GCPD, V=6.309 according to Simbad, V=6.307 according to Tycho2. Its
$B-V$ colour index is slightly more uncertain : $B-V=0.635$ according to the GCPD, $B-V=0.604$ according
to Simbad, $B-V=0.646$ according to Tycho2, but suggests however a higher temperature than the Sun. 
This is confirmed by several recent estimates of Teff available in the literature and spanning values 
from 5719K (Chen et al \cite{CNZZB00}) to 5833K (Blackwell \& Lynas-Gray \cite{BLG98}) with a mean 
value of 5801K. This large dispersion is a good illustration of the lack of a common temperature 
scale, even for bright nearby stars.
HD168009 is also part of a catalogue of high precision near infrared photometry by Kidger \& Martin-Luis 
(\cite{KML03}) who give J=5.133, H=4.840, K=4.783, values which differ by less than 0.005 mag
from those measured by Alonso et al (\cite{AAM98}). A colour index V-K=1.512 leads to Teff=5730 K
with the relation established by Alonso et al (\cite{AAMR96}).

The abundance of several elements (O, Na, Mg, Al, Si, K, Ca, Ti, V, Cr, Ni, Ba) have been measured
by Chen et al (\cite{CNZZB00}) to be solar within the error bars. Ba, Eu and Sr abundances have
also been measured by Mashonkina \& Gehren (\cite{mash01}), leading to the same conclusion.

\subsection{HD010307}

Like HD168009, HD010307 seems to be hotter and more luminous than the Sun. Allende Prieto et
al (\cite{all99}) quote a mass of $0.94M_\odot$ and an astrometric gravity of logg=4.29, in very good 
agreement with the averaged spectroscopic gravity of logg=4.26 quoted in Table \ref{t:general_table}.
HD010307 is in fact a binary system which was resolved by Henry et al (\cite{hen92}), the low mass 
companion being 1000 times fainter. According to a detailed analysis of Hipparcos data by Martin 
et al (\cite{mar98}) the system has a total mass of $(0.931 \pm 0.178) M_\odot$,
a primary mass of $(0.795 \pm 0.159) M_\odot$ and a secondary mass of $(0.136 \pm 0.053) M_\odot$.
At a distance of
12.6 pc (12.4 pc when the binarity is considered), it is the nearest of our Top Ten solar analogs. 
It is a well studied star with many measurements
in good agreement of its apparent visual magnitude and $B-V$ colour, and it is part of the catalogue of the least 
variable stars compiled by Adelman (\cite{ade01}) despite its binarity. We notice that the temperature given
by Chen et al
is similar to that of the Sun but significantly lower than given by other authors, as was also the
case for  HD168009.
Hardorp (\cite{H78}) mentioned HD010307 with the comment "some aborption features appreciably weaker than 
solar" which is in agreement with a higher temperature.

Cayrel de Strobel (\cite{cay96}) and Fesenko (\cite{F94}) have included HD010307 in their list of
solar analogs but not with a high rank. It was only mentioned to "deserve 
study" in the conclusions of the Lowell Workshop. 

\subsection{HD089269}

Very few papers mentioning  HD089269 are available in the literature. It was never recorded as a 
solar analog despite its colour index $B-V=0.654$ being similar to that of the Sun. Its visual magnitude 
V=6.633 combined to its trigonometric parallax $\pi=48.45 \pm 0.85$ mas leads to an absolute magnitude 
$M_{\rm V}=5.06$ indicating that it is less massive than the Sun. Mishenina et al (\cite{Mish03}) find it
100K colder than the Sun and significantly more metal poor. The good score obtained with TGMET
indicates that, globally,  the combined effects of temperature and 
iron abundance may give similar absorption features as in the Sun.

\subsection{HD047309}

HD047309 is unknown as a solar analog. The only data are Str\"omgren photometry, 
and data from Hipparcos and Tycho2. Contrary to HD089269, it seems to be slightly hotter, more 
luminous and metal-rich. It is also the most distant of our sample at 42.4 pc.
	
\subsection{HD095128}
	
We come back to well a known star with HD095128 (47 UMa) known to have two giant planets 
orbiting around it. Consequently it has been very well studied. Its temperature is 
significantly higher than that of the Sun. Again the temperature given by Chen et al is the lowest 
of the list
with 5731K whereas Santos et al (\cite{SIMRU03}) find the highest temperature, with 5925K. The
2 colour indexes $B-V=0.606$ and $b-y=0.391$ confirm a higher temperature than the Sun. Despite
the large dispersion in Teff, the dispersion in [Fe/H] is low, with an
average exactly solar.   HD095128 is more luminous than the Sun.
Allende Prieto et
al (\cite{all99}) quote a mass of $0.96M_\odot$ and an astrometric gravity of logg=4.23, in good 
agreement with the averaged spectroscopic gravity of logg=4.28 quoted in Table \ref{t:general_table}.
It shows that HD095128 has already evolved from the main sequence.

\subsection{HD042618}

The only mention of HD042618 is by Fesenko (\cite{F94}) in his solar type star study. 
The 2 available 
estimates of its temperature differ by 120K, but $B-V = 0.632$  suggests that the hottest one,
5775K,
is the most likely. The cold temperature scale adopted by Reddy et al (\cite{red02}) implies a
low metallicity of [Fe/H]=-0.16, which might be closer to the solar one in fact. It is not very clear 
however, because its absolute magnitude $M_{\rm V}=5.05$ combined with a solar temperature 
would not be compatible with being on the solar composition ZAMS.

\subsection{HD071148}	

Like HD042618, HD071148 was recorded by Fesenko (\cite{F94}) but did not receive
much attention as a solar analog. Its effective temperature
is subject to a controversy between partisans of a low temperature scale (Reddy et al \cite{red02}
and Chen et al \cite{chen03}) giving a temperature of about 5710K and Mishenina et al (\cite{Mish03})
giving a temperature 140K higher. Again we find that the colour indexes indicate that this star is
probably slightly hotter than the Sun. The 3 authors agree on the fact that its metallicity is 
nearly solar. 

Interestingly, the radial velocity of HD071148 has been monitored during several years by Naef 
et al (\cite{naef03}) who found that it is constant within 10 m\,s$^{-1}$, ruling out the presence 
of a low mass companion.
	
\subsection{HD186427}
HD186427 (16 Cyg B) is one of the best solar twin candidates of the 
Lowell Fall Workshop and also a planet-host star.
Its UV spectrum was qualified as "indistinguishable from solar" by Hardorp (\cite{H78}), a similarity 
confirmed by Fernley et al (\cite{fer96}). There is a remarkable agreement between the 10 authors 
who have estimated its temperature, colder than the Sun by only $\sim$20K. Only Laws \& Gonzalez
(\cite{LG01}) and  Gonzalez (\cite{G98}) used a significantly lower temperature scale. HD186427 differs
from the Sun mainly by its higher metal content, and by a higher luminosity. 
 
\subsection{HD010145}
HD010145 has never been mentioned as a solar analog and was little studied before. The recent 
spectroscopic analysis performed by Mishenina et al (\cite{Mish03}) shows that it is colder than the
Sun by 100K, but with a similar gravity and metal content. It is the coldest of our Top Ten together with
HD089269, the latter one having bluer colour index consistent with its lower metallicity. HD010145 has a 
lower luminosity than the Sun, indicating either a lower age or a lower mass.

\section{Li content, activity, ages}
\label{s:Li}
In this section we compare qualitatively the Li content of our Top Ten solar analogs with that of the Sun. 
The solar photosphere is known to be highly depleted in Li, as is the case for many solar type stars. This depletion is 
however subject to various interpretations, involving rotation, convection, or the presence
of a planetary system. The correlation of age with Li depletion has also been often discussed but not fully
established. 

\begin{figure}[hbtp]  
\centering  
      \includegraphics[width=4.3cm]{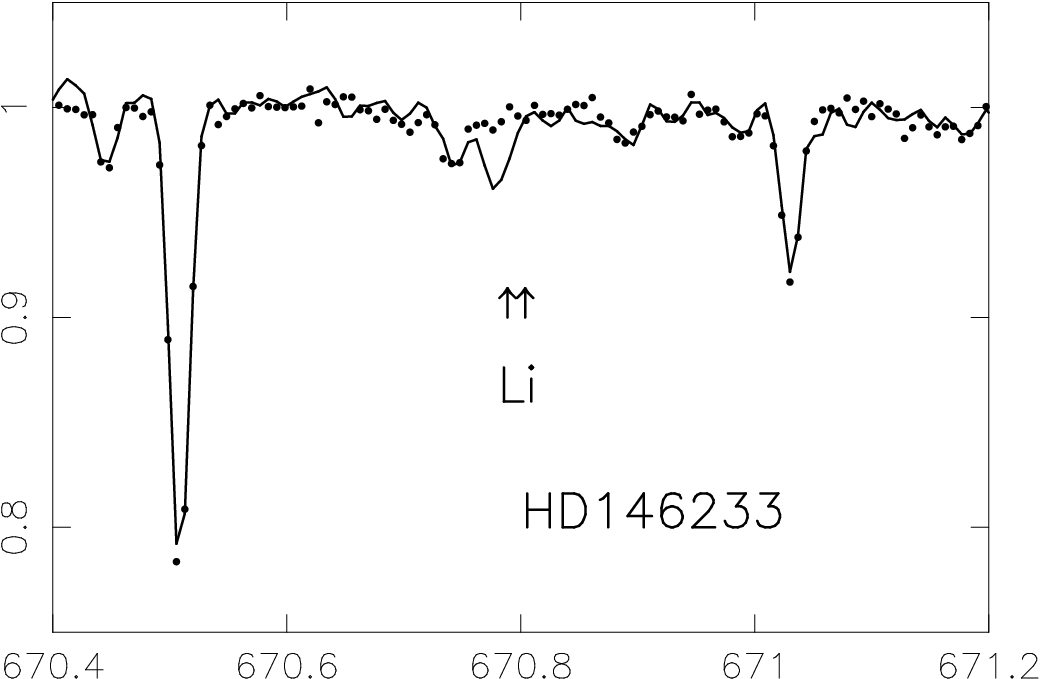}
      \includegraphics[width=4.3cm]{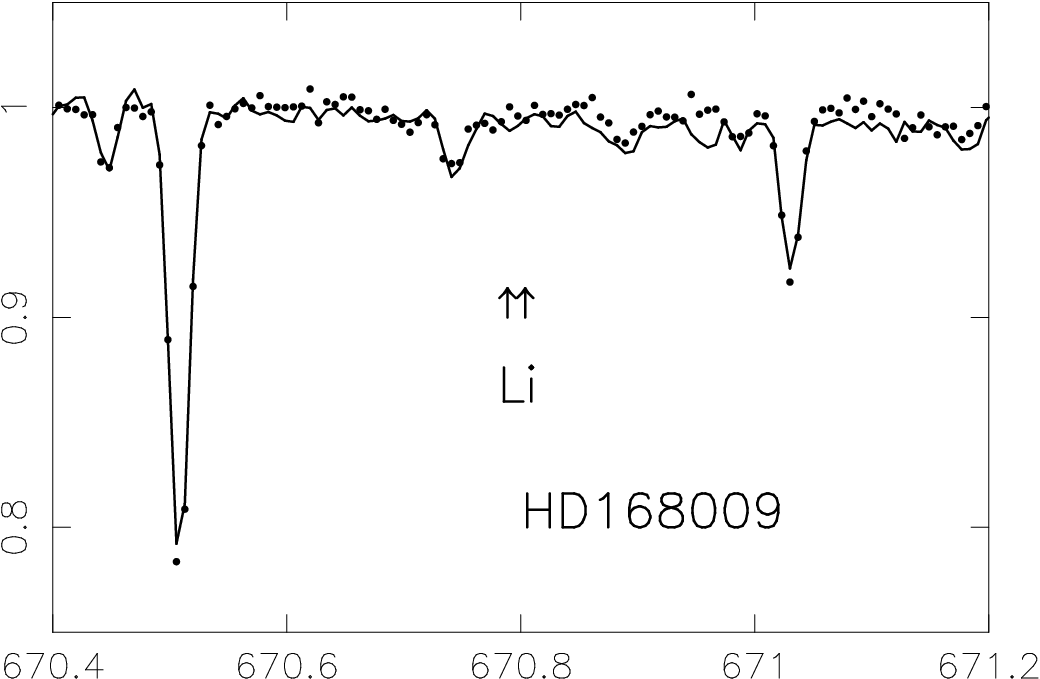}
      \includegraphics[width=4.3cm]{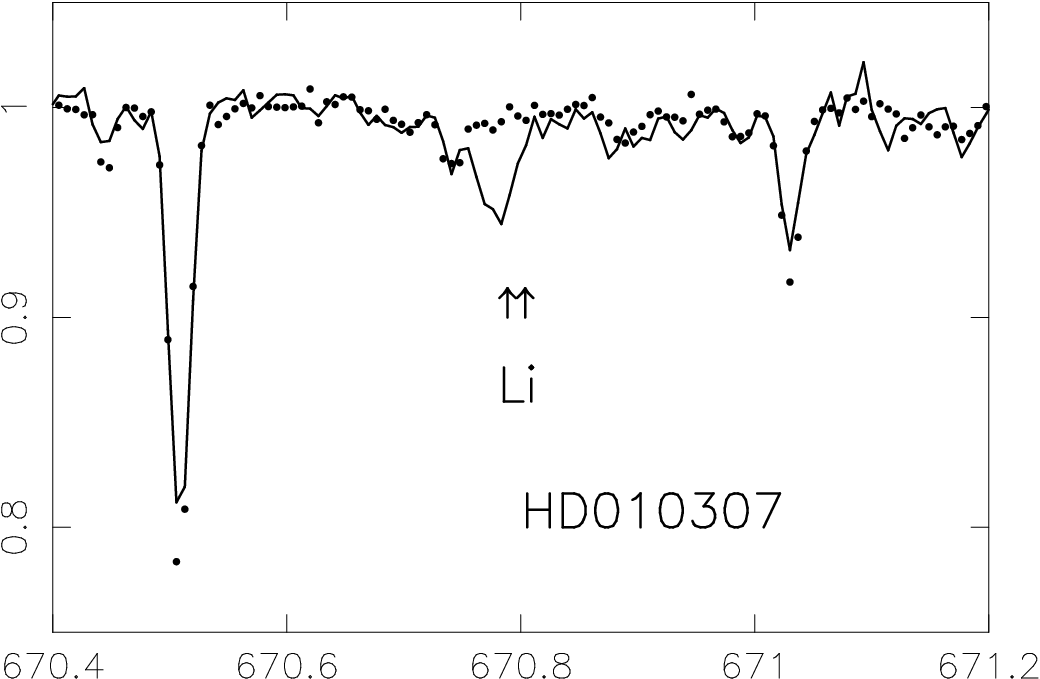}
      \includegraphics[width=4.3cm]{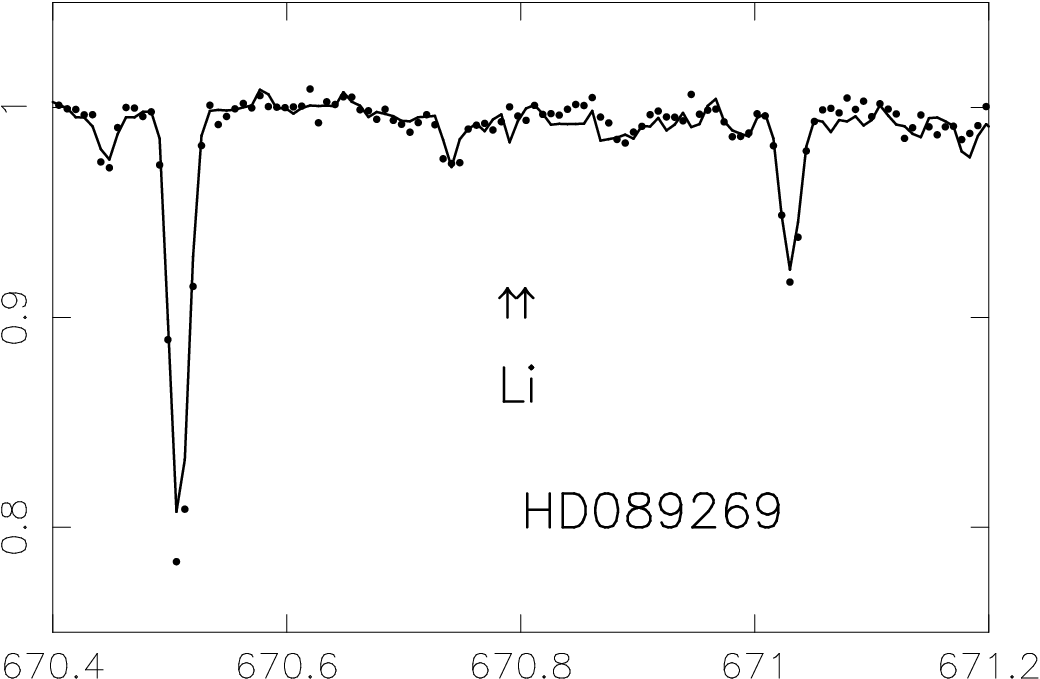}
      \includegraphics[width=4.3cm]{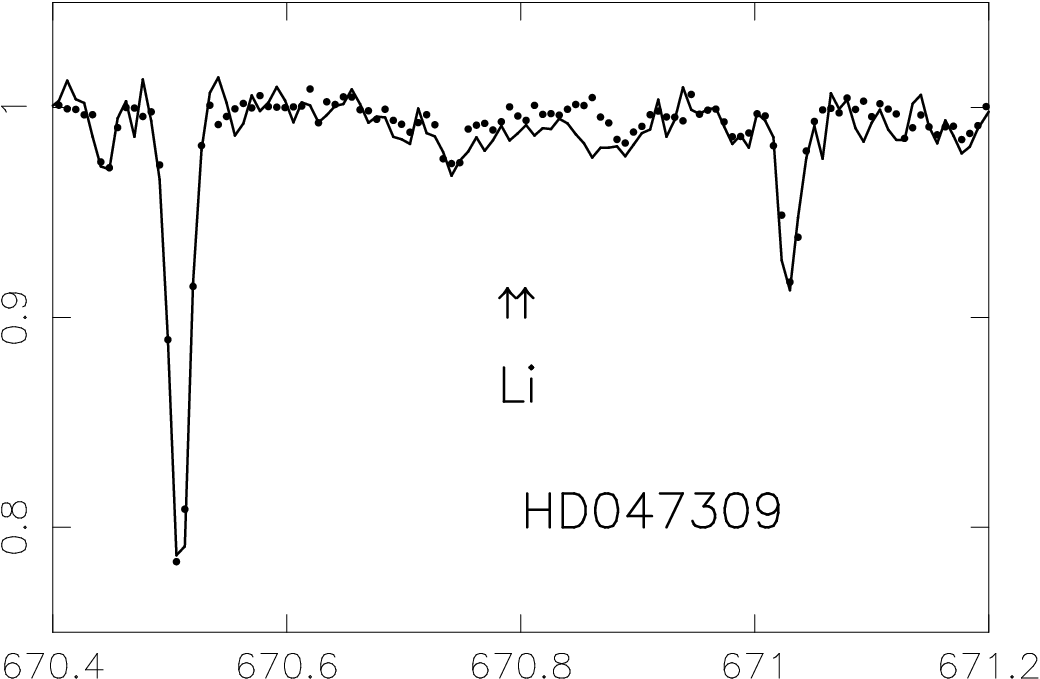}
      \includegraphics[width=4.3cm]{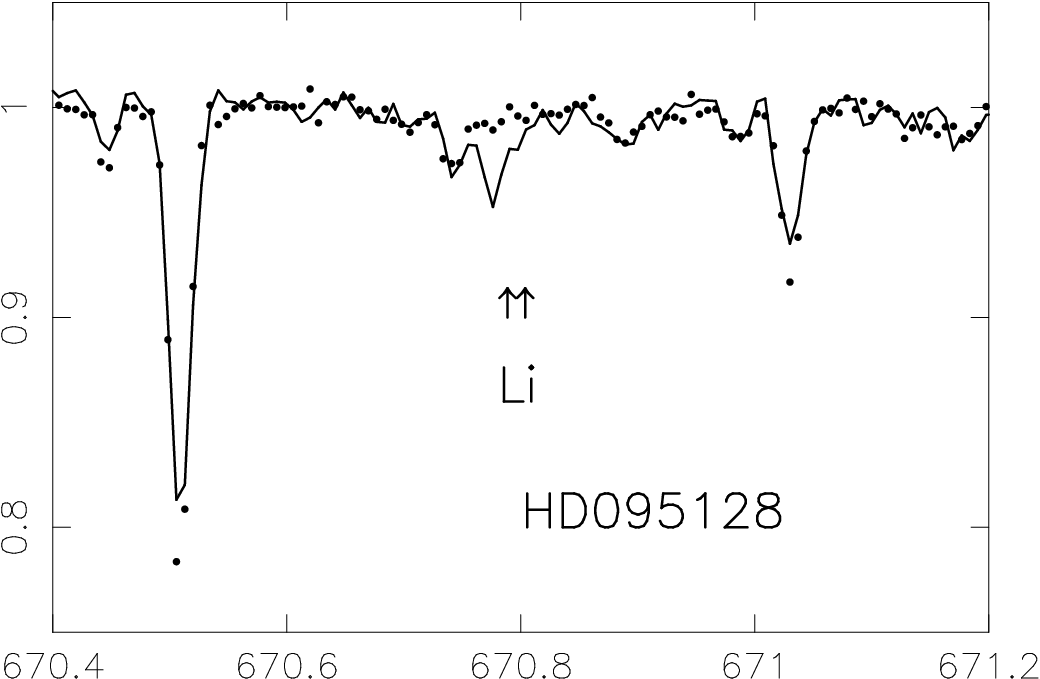}
      \includegraphics[width=4.3cm]{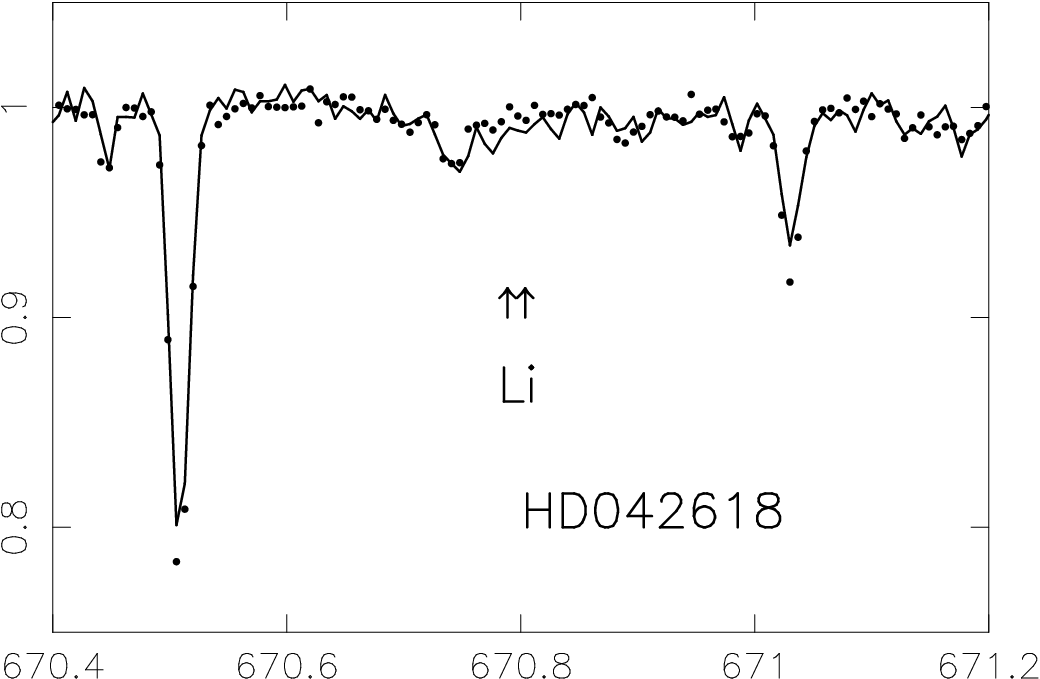}
      \includegraphics[width=4.3cm]{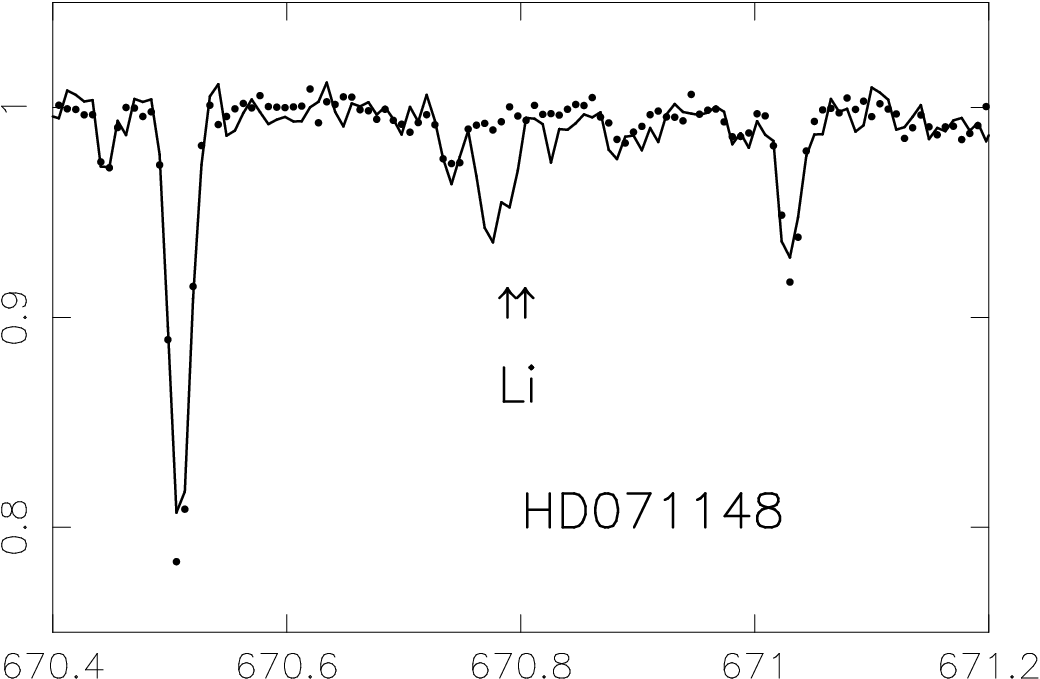}
      \includegraphics[width=4.3cm]{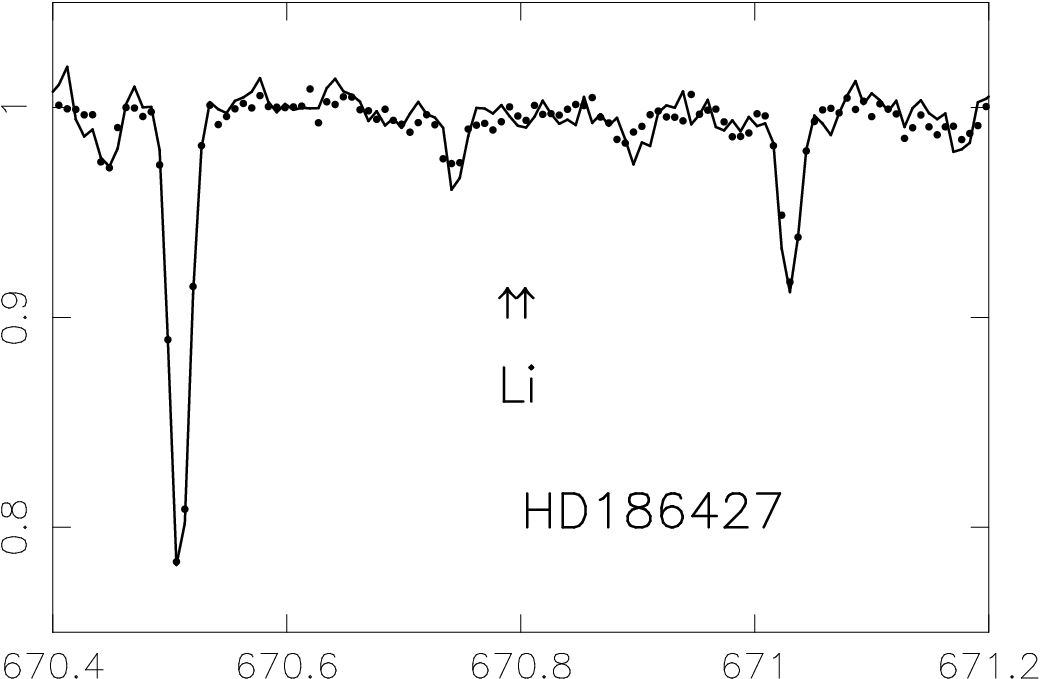}
      \includegraphics[width=4.3cm]{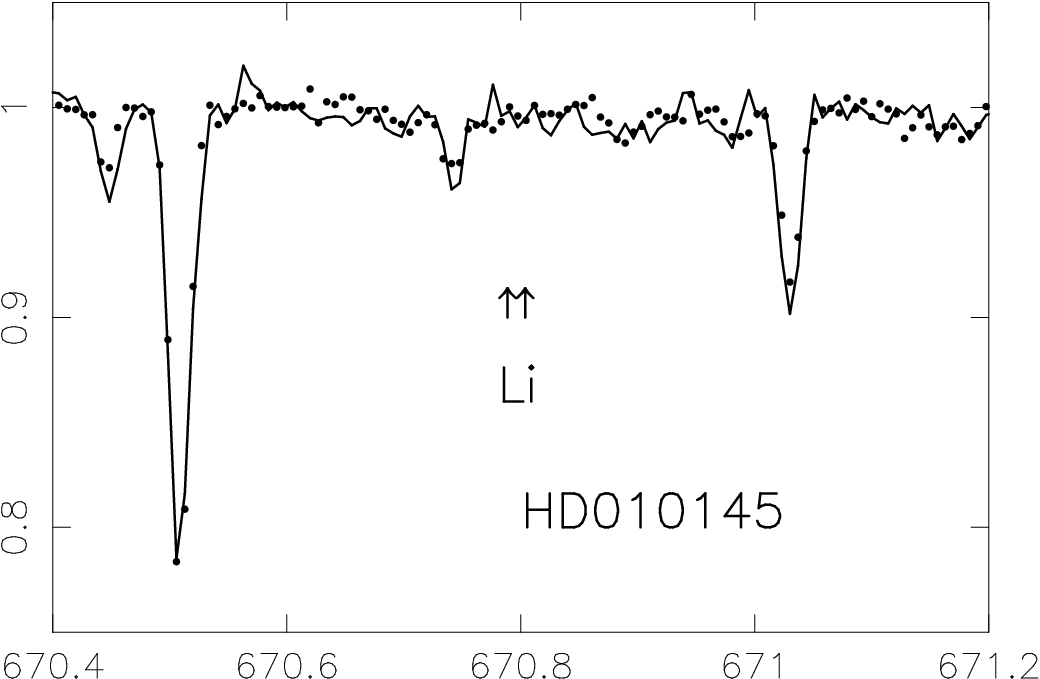}
      \caption{Comparison between the Li I 670.7 nm region of the Sun (spectrum 00903, dotted line) and
      of each Top Ten solar analog} 
\label{f:Li} 
\end{figure}

The $^7$Li doublet resonance lines at 670.78 nm and 670.79 nm are well placed in 
the middle of the 66th order of the ELODIE spectra. In Fig. \ref{f:Li}, for each of our Top Ten, the Li
 region was superposed on
the solar spectrum, showing that 6 of them are depleted in Li like the Sun whereas the 4 others show
a pronounced Li feature. The most pronounced feature concerns HD071148 and HD010307. A weaker feature is 
also present in the spectra of HD095128 and HD146233. 

More clearly than the Li content, the chromospheric activity of a solar type star is directly connected to its
age. Thus it would have been extremely interesting to look at the chromospheric activity revealed by the central depth
of the Ca II H and K lines, and of the Ca II triplet lines at 852 nm. Unfortunatly the NIR Ca II triplet is
not in the spectral range of ELODIE and the H and K lines appear on the border of the 2nd and 3rd orders 
which are underilluminated. The core of the H$_\alpha$ line is also an indicator of chromospheric activity,
but we were not able to detect any significant difference of depth, even for the 4 stars
having a higher Li content. According to Soderblom (\cite{sod85}), HD071148, HD010307 and HD095128 show
CaII emission strengths and rotation similar to  the Sun, suggesting a weak activity. Hall \& 
Lockwood (\cite{HL00}) found an  activity cycle in HD146233 with an amplitude slightly greater than that of
 the Sun.
Thus these 4 stars presenting a pronounced Li feature do not seem to be enormously more active than the Sun.
Two other analogs are part of Soderblom's study, HD010145 and HD168009; they do not exhibit evidence of
a higher activity than the Sun.

Several of our Top Ten had their age estimated by Ibukiyama \& Arimoto (\cite{ibu02}) : 7.02 Gyr for HD168009,
7.32 Gyr for HD010307, 6.92 Gyr for HD095128, 6.65 Gyr for HD071148. According to Cayrel de Strobel \& Friel 
(\cite{cay98}), HD146233 has an age of 6 Gyr. Thus one can find stars older than the Sun which 
have a higher Li content.

\section{Solar analogs selected by other methods}
\label{s:other}
The ELODIE library includes the spectra of several stars which were considered in previous studies 
as high rank solar analogs. We will focus here on HD217014 (51 Peg), HD028099 (Hy VB 64) and HD186408 
(16 Cyg A). 

With
our purely differential method HD186408 is within the 20 best solar analogs among our list of
208 dwarfs. A dozen recent determinations of its atmospheric parameters are available and
give on average Teff=5780K, logg=4.26, [Fe/H]=+0.07. Its visual absolute magnitude is $M_{\rm V}=4.29$, to be 
compared to $M_{\rm V}=4.56$ for its companion HD186427 and $M_{\rm V \odot}=4.82$ showing that HD186408 is more massive 
and evolved. 

HD217014 and HD028099 obtained a lower score in the TGMET output despite their similar effective 
temperature and gravity to the Sun. These 2 stars have also been very
much studied, especially HD217014 because of its planet. The latest determination of its atmospheric
parameters led to Teff=5805K, logg=4.51, [Fe/H]=+0.21 (Santos et al \cite{ SIMRU03}). Thus
stronger metallic lines explain that it is not a good spectroscopic analog of the Sun. HD028099 was
found to have Teff=5800K, logg=4.40, [Fe/H]=+0.10 by Paulson et al (\cite{paul03}). This star is not
considerably more metal rich than the Sun, but it is known to be younger and to have a high chromospheric 
activity. Looking closely at its spectrum shows that its lines are not as deep as those in the solar spectrum.

We have also searched for photometric solar analogs in the ELODIE library by selecting stars 
having the same colours and absolute 
magnitude as the Sun. Four stars fall in the narrow range $4.6 < M_{\rm V} < 5.0, 0.63 < B-V < 0.68,
0.16 < U-B < 0.23, 0.40 < b-y < 0.42$ : HD001835 (BE Cet), HD076151, HD146233, HD159222. 
HD076151 in not 
in our Top Ten but it is at the 14th position. It is a well 
studied star, with recent determinations of its atmospheric parameters  in good agreement, 
giving on average Teff=5774K, logg=4.39, [Fe/H]=+0.06. It was also one of the good solar twin candidates
discussed by Cayrel de Strobel (\cite{cay96}). However, this star is younger than the Sun 
(3.04 Gyr estimated by Ibukiyama \& Arimoto \cite{ibu02}), with a stronger activity and 
faster rotation (Pizzolato et al \cite{piz03})
and presents a pronounced Li feature. In combination with a higher metal abundance, this may explain
why it was not at higher rank in the TGMET output. HD001835 is a variable star of the BY Dra type
which renders its photometry suspect. It was also discussed by Cayrel de 
Strobel (\cite{cay96}) as a good solar analog for the temperature and gravity but not for Li, 
chromospheric activity and age. Its young age is assessed by its membership of the Hyades moving group
and its high activity is confirmed by Pizzolato et al (\cite{piz03}). It is also more
metal-rich than the Sun ([Fe/H]=+0.13 estimated by Mishenina et al  \cite{Mish03}) and its distance
to the solar spectra computed by TGMET is very large. Its activity is clearly seen in its H$_\alpha$ 
line, which is shallower than in the Sun. HD159222 the 11th star in the TGMET output. It is thus a good
photometric and spectroscopic solar analog. Moreover its age is also very similar (4.56 Gyr estimated 
by Ibukiyama \& Arimoto \cite{ibu02}). Several determinations of Teff are available for this star,
showing an impressive dispersion : 5708 K by Blackwell \& Lynas-Gray (\cite{BLG98}), 5770 K by 
Alonso et al (\cite{AAM96}), 5834 K by Mishenina et al  (\cite{Mish03}), 5852 K by di Benedetto 
(\cite{dB98}).

We have also observed that many spectra were polluted by telluric lines in the red orders, and that order 39
including the MgI triplet was a powerful discriminator of solar resemblance. We have thus performed the TGMET
comparison of the 8 solar spectra with the library using only this order. This has greatly modified the order of our 
list, HD146233 keeping however its highest rank. The five closest stars are part of the Top Ten list : HD146233,
HD047309, HD168009, HD042618 and HD186427. The 4 stars of the Top Ten list with the lowest temperatures 
(HD010145 and HD089269) and 
the highest temperatures (HD010307 and HD095128) are pushed away. Five new stars appear in the 10 closest
solar analogs : HD195034, HD159222 (also a good photometric analog), HD187123, HD186104, HD005294. 
The photometric analog HD076151 is at the 11th position.

The parameters of the solar analogs discussed in this section are listed in Table \ref{t:table_other}. 
All of them, except HD005294, are more metal-rich than the Sun. 

	\begin{table*}
	\caption{
	\label{t:table_other} 
	Fondamental data for solar analogs mentioned in Sect \ref{s:other}. Only the most recent 
	determination of atmospheric parameters is listed. The second colomn indicates how the star was selected as
	a solar analog (P : photometry, L : Lowell workshop, Mg : TGMET on the Mg I triplet region) }
	\begin{center}
	\begin{tabular}{ c c c c c c c c c l }
	\hline
	\hline
	Star name & method & $B-V$ & $b-y$& $U-B$ & $M_{\rm V}$ & Teff & log g & [Fe/H] & source              \\
	\hline
	HD186408 & L  & 0.645 & 0.410 & 0.187 & 4.29 & 5803 & 4.20 & +0.02 & Mishenina et al  (\cite{Mish03})\\
	HD217014 & L  & 0.665 & 0.416 & 0.224 & 4.53 & 5805 & 4.51 & +0.21 & Santos et al (\cite{SIMRU03})\\
	HD028099 & L  & 0.660 & 0.411 &       & 4.75 & 5800 & 4.40 & +0.10 & Paulson et al (\cite{paul03})\\
	HD076151 & P  & 0.662 & 0.413 & 0.217 & 4.83 & 5776 & 4.40 & +0.05 & Mishenina et al  (\cite{Mish03})\\
	HD001835 & P  & 0.659 & 0.409 & 0.226 & 4.84 & 5790 & 4.50 & +0.13 & Mishenina et al  (\cite{Mish03})\\
	HD005294 & Mg & 0.650 & 0.401 & 0.174 & 5.03 & 5779 & 4.10 & -0.17 & Mishenina et al  (\cite{Mish03})\\
	HD159222 & P+Mg & 0.637 & 0.406 & 0.172 & 4.67 & 5834 & 4.30 & +0.06 & Mishenina et al  (\cite{Mish03})\\
	HD186104 & Mg & 0.631 & 0.412 &       & 4.62 & 5753 & 4.20 & +0.05 & Mishenina et al  (\cite{Mish03})\\
	HD187123 & Mg & 0.619 & 0.405 &       & 4.43 & 5855 & 4.48 & +0.14 & Santos et al (\cite{SIMRU03})\\
	HD195034 & Mg & 0.610 & 0.408 &       & 4.84 &   &   &   &  \\
	\hline
	\end{tabular}
	\end{center}
	\end{table*}

\section{Discussion}

 It is interesting to note that, despite the great similitude of the optical spectrum of our 
 Top Ten solar analogs
to that of the Sun, their atmospheric parameters can differ significantly. Effective temperatures span
$\pm$100K on both sides of the solar value, logg values span the interval [4.09 ; 4.58] and [Fe/H]
span the interval [-0.23 ; +0.11]. Several interpretations can explain this dispersion. On the one
hand, authors do not use the same temperature scale and model atmospheres. Temperature scales can differ
by more than 150K as we have seen. It is very important that authors agree on temperatures because 
this parameter has a strong impact on abundance determinations. Uncertainties which affect the stellar parameter 
determinations
have been discussed by Asplund (\cite{asp03}) to be of the order of 50K to 100K in Teff, 
0.2 dex for logg and 0.1 dex for [Fe/H]. On the other hand, we cannot expect finding a perfect
twin having all its parameters exactly solar, especially in an incomplete sample. Finally, temperature and
metallicity have contrary effects on the overall spectrum which may compensate in some cases (ex HD089269 
or HD187123). It is also possible that other effects act on the spectra. Observing conditions and telluric
lines are the most obvious, but intrinsic stellar properties also have an influence on the spectrum. We
have mentioned chromospheric activity and rotation, but the abundance of other elements than iron,
turbulent motions, spots on the stellar surface may be different than in the Sun. For instance, when using 
TGMET only in the MgI triplet region, the Mg abundance of the star may have a strong weight.

\begin{figure}[hbtp]  
\centering  
      \includegraphics[width=4.2cm]{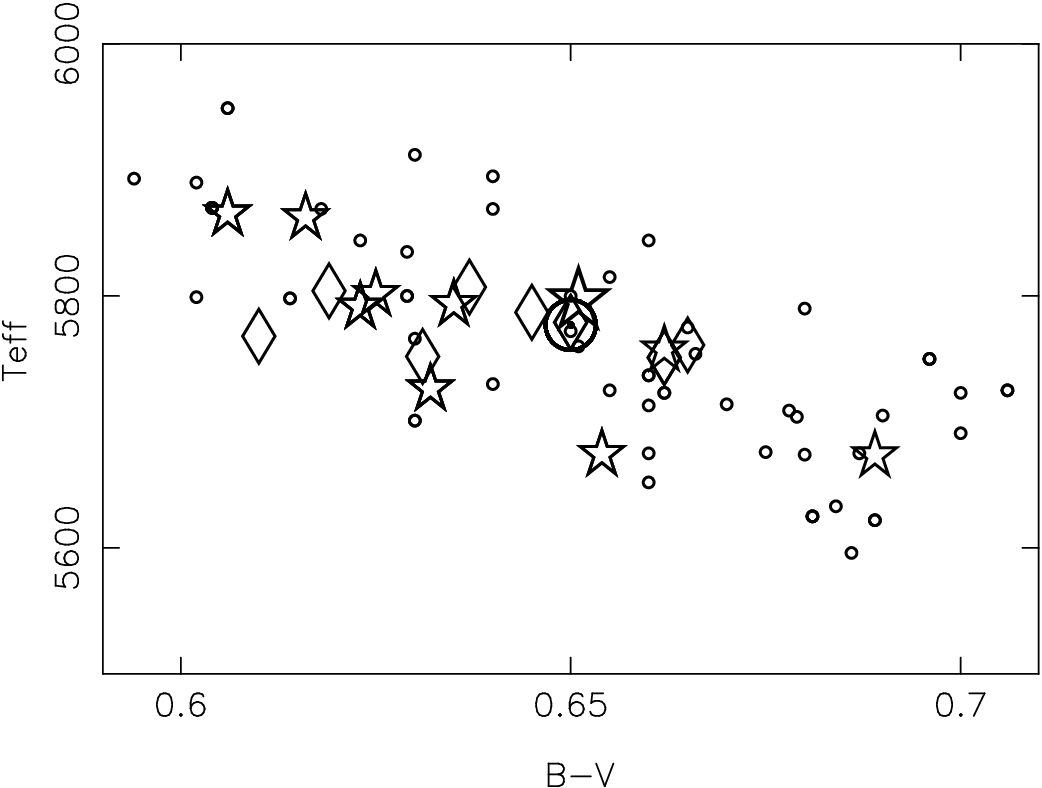}
      \includegraphics[width=4.2cm]{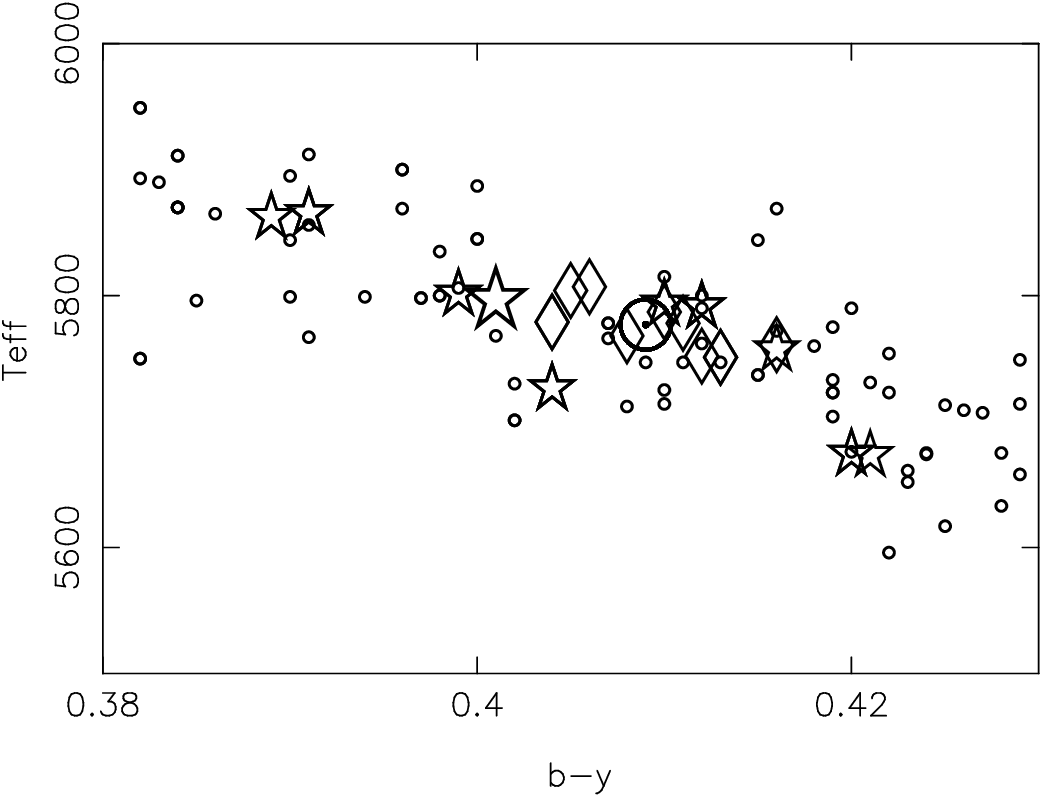}
      \includegraphics[width=4.2cm]{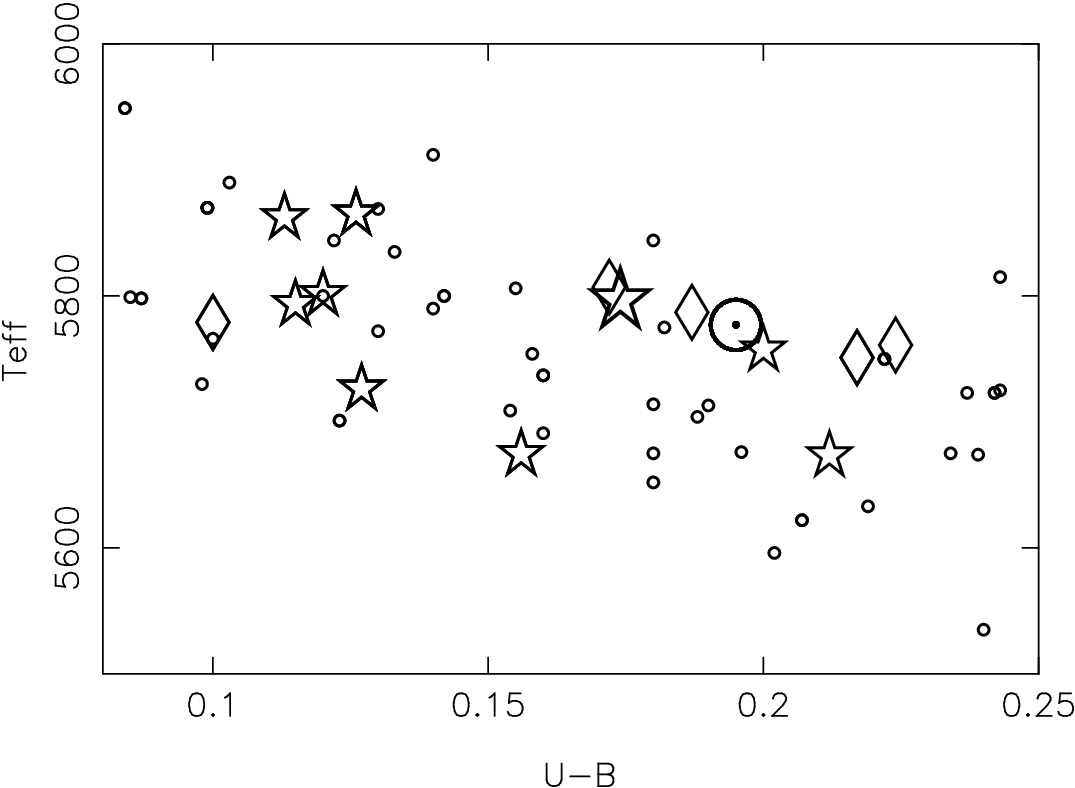}
      \includegraphics[width=4.2cm]{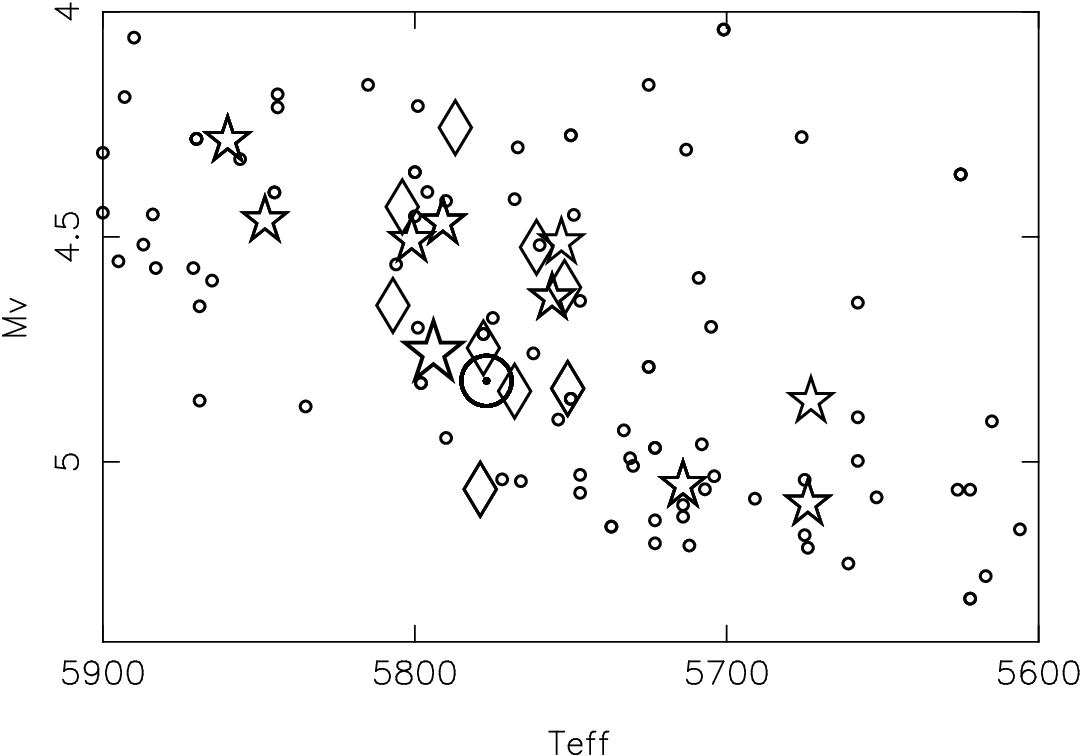}
      \caption{Distribution of the ELODIE library (small dots), the Top Ten solar analogs (stars), other
      solar analogs (Sect. \ref{s:other}, diamonds)  and the Sun in  Teff vs colour and HR diagrams. The
      solar twin HD146233 is shown with a larger symbol.} 
\label{f:teff_colour} 
\end{figure}

Colour indexes of the Top Ten solar analogs 
also span intervals as large as 0.073 in $B-V$, 0.032 in $b-y$, 0.099 in $U-B$, and absolute magnitudes 
range from $M_{\rm v}=4.31$ to $M_{\rm v}=5.06$. Fig. \ref{f:teff_colour} represents their 
distribution in Teff vs colour 
and HR diagrams together with the other analogs discussed in Sect. \ref{s:other} and the rest of
the ELODIE library, restricted to $-0.25 <$ [Fe/H] $< +0.15$. The Str\"omgren index $b-y$ is clearly the 
one 
which presents the lowest dispersion in its relation with Teff. Like atmospheric parameters, 
magnitudes and colours
are affected by uncertainties and a lack of homogeneity. Absolute magnitudes are computed from excellent
parallaxes but averages of old and inhomogeneous apparent magnitudes. Tycho2 (H{\o}g et al \cite{HOG00})
is a recent photometric catalogue of good quality but its B and V passbands do not correspond to the
Johnson standard system and transformations, also affected by calibration uncertainties, have to be used. 
A small fraction of the dispersion may also be due to interstellar absorption, even if our targets are 
closer than 50 pc. But we interpret the larger part of the dispersion to mean that our incomplete 
sample of 208 G dwarfs includes stars of various ages and states of evolution resulting in a variety of
astrophysical properties, because the ELODIE library was built to represent the stellar content in the 
solar neighbourhood, not solar analogs. 

We have seen in previous sections that good photometric analogs are not always good spectroscopic analogs. 
HD001835 is a good example of a star having similar colours and absolute magnitude as
the Sun but which is considerably different in respect to its age, activity and metal content. Thus photometry
is not able to discriminate between these effects whereas high resolution spectroscopy can. In
contrast the direct comparison of high resolution spectra used alone classify as solar analogs stars
having a large range of atmospheric parameters. We conclude that   a good strategy to find 
other solar twins than HD146233, and perhaps better ones, would be to select photometric analogs in large
catalogues, then select with Hipparcos those that have a similar absolute magnitude to the Sun, 
then submit their  high resolution spectrum to the TGMET comparison. We have scanned the GCPD, Str\"omgren 
and Hipparcos catalogues with the drastic criterion used in Sect. \ref{s:other} to identify photometric
analogs and found only 27 candidates, 4 being already in the ELODIE library, 15 others observable with
ELODIE. We plan to observe them soon in order to complete this work.

\section{Conclusion}

We have presented the 10 stars of the ELODIE library which exhibit the closest optical spectrum
to the Sun at a resolution of 42000. They have colours,
absolute magnitudes, atmospheric parameters and Li content which span a range of values larger than expected.
It is surprising that a star like HD089269, colder, more metal poor and less luminous than the Sun
is at the 4th position, whereas HD076151, having similar colours, absolute magnitude and atmospheric 
parameters is only at the 14th position. Activity may play an important role in discrimination. We have
shown for instance that the good photometric analog HD001835 was a very bad spectroscopic analog
because of its high activity. One also has to take into account, when comparing colours, absolute magnitudes
and atmospheric parameters to those of the Sun, that these quantities are affected by significant 
uncertainties. Effective temperatures are particularly in question, with determinations for the same star
differing by nearly 200K in some cases.  Our method consisting in measuring distances
between spectra is powerful but it is also affected by uncertainties due to 
observing conditions, especially the pollution by telluric lines, which may perturb the order of the 
classification.

Among our Top Ten, several stars have never been mentioned before as solar analogs and have 
been very little studied.
They are good candidates for planet hunting, especially HD047309 which is slightly more metal rich
than the Sun. Two of our solar analogs, HD095128 and HD186427, are already known to have planets. HD159222
and HD076151 are also good candidates because they are good spectroscopic analogs (in the Top 15) and good 
photometric analogs.

The conclusion of this work is that none of the methods to search for solar twins is satisfactory when used 
by itself. The methods that have been already used are the comparison of colour indexes, of absolute magnitudes, 
of UV spectral energy distributions, of atmospheric parameters and of high resolution optical spectra.
All these methods are affected by uncertainties and none of them is able to describe sufficiently all
the stellar properties. Combining them is the best way to minimize their drawbacks, uncertainties and
insufficiencies. Finally HD146233 is the only star in the ELODIE library which merits the title of solar 
twin because it has passed the filter of all methods. It is not however a perfect twin and differs
from the Sun by its higher Li content, slightly higher age (6 Gyr against 4.6 Gyr for the Sun) 
and higher luminosity ($M_{\rm V}=4.77$ against $M_{\rm V \odot}=4.82$).

\begin{acknowledgements}
This  research  has made  use  of  the  SIMBAD and VizieR,
operated at CDS, Strasbourg,  France.
We warmly thank G. Cayrel de Strobel who has suggested and discussed this work and
J.-C. Mermilliod who gave us
the averaged version of the GCPD. We are grateful to the observers on ELODIE who made 
their observations available for the ELODIE library, specially the Corot group leaded by C. Catala and 
the extrasolar planet group leaded by M. Mayor who are the authors of some of the spectra used in this paper.
AT aknowledges financial support from the Student-Staff council of the St-Andrews University.
\end{acknowledgements}
%
%
%

\landscape
	\begin{table}[!hb]
	\caption{
	\label{t:general_table} 
	Fondamental data for our Top Ten solar analogs. Our adopted atmospheric parameters in
	bold characters are the mean values from the literature. The second column is the averaged TGMET
	distance of the corresponding star to the 8 solar spectra.}
	\begin{center}
	\begin{tabular}{ c c c c c c c c c r l }
	\hline
	\hline
	Star name & TGMET &  $B-V$ & b-y& U-B & distance & Absolute & Teff & log g & [Fe/H] & source              \\
	          & score &      &    &     & in pc    & magnitude&      &       &        &(Teff, logg, [Fe/H]) \\
	\hline
	Sun & & 0.64-0.66  & 0.40-0.41 & 0.17-0.20 & 0 & 4.80-4.84 & 5777 & 4.44 & 0.00 & - \\
	\hline
	HD146233 & 2.019 & 0.651 & 0.401 & 0.174 & 14.0 & 4.77 & 5799 & 4.40 & +0.01 & Mishenina et al  (\cite{Mish03})\\
	&&&&&&&  5789 & 4.49 & +0.05 & Porto de Mello \& da Silva (\cite{PMDS97}) \\
	&&&&&&&  \bf{5794} &\bf{4.44} & \bf{+0.03}\\
	\hline
	HD168009 & 2.195 & 0.635 & 0.410 & 0.115 & 22.7 & 4.52 & 5826 & 4.10 & -0.01 & Mishenina et al  (\cite{Mish03})\\
	&&&&&&& 5719 & 4.08 & -0.07 & Chen et al (\cite{CNZZB00})\\
	&&&&&&& 5826 & - & - & di Benedetto (\cite{dB98})\\
	&&&&&&& 5833 & - & - & Blackwell \& Lynas-Gray (\cite{BLG98})\\
	&&&&&&& 5781 & -& - & Alonso et al (\cite{AAM96})\\
	&&&&&&& \bf{5801}  &  \bf{4.09}  & \bf{-0.04}  \\
	\hline
	HD010307 & 2.516 & 0.616&  0.389 & 0.113 & 12.6 & 4.45 & 5881 & 4.30 & +0.02 & Mishenina et al  (\cite{Mish03})\\
	&&&&&&& 5776 & 4.13 & -0.05 & Chen et al (\cite{CNZZB00})\\
	&&&&&&& 5825 & 4.33 & -0.04 & Gratton et al (\cite{GCC96}) \\
	&&&&&&& 5874 & - & - & Alonso et al (\cite{AAM96})\\
	&&&&&&& 5898 & 4.31 & -0.02 & Edvardsson et al (\cite{EAGLNT93})\\
	&&&&&&& \bf{5848}  &  \bf{4.26}  &  \bf{-0.02}  \\
	\hline
	HD089269 & 2.602 & 0.654& 0.420 & 0.156 & 20.6 & 5.06 & 5674 & 4.40 & -0.23 & Mishenina et al (\cite{Mish03})\\
	&&&&&&& \bf{5674} &  \bf{4.40} &\bf{-0.23} \\
	\hline
	HD047309 & 2.628& 0.623&  0.412 & - & 42.4 & 4.47 & 5791 & - & - & Kovtyukh et al (\cite{KSBG03})\\ 
	&&&&&&& 5791 & 4.40 & +0.11 & Mishenina priv. com.\\
	&&&&&&& \bf{5791} &  \bf{4.40}& \bf{+0.11} \\
	\hline
	HD095128 & 2.758& 0.606 &  0.391 & 0.126 & 14.1 & 4.31 & 5887 & 4.30 & +0.01 & Mishenina et al  (\cite{Mish03})\\
	&&&&&&& 5861 & 4.29 & +0.05 & Laws et al (\cite{Laws03})\\
	&&&&&&& 5925 & 4.45 & +0.05 & Santos et al (\cite{SIMRU03})\\
	&&&&&&& 5788 & 4.31 & +0.03 & Zhao et al (\cite{ZCQL02})\\
	&&&&&&& 5731 & 4.16 & -0.12 & Chen et al (\cite{CNZZB00})\\
	&&&&&&& 5892 & 4.27 & 0.00 & Zhao \& Gehren (\cite{ZG00})\\
	&&&&&&& 5892 & 4.27 & 0.00 & Fuhrmann et al (\cite{FPB98})\\
	&&&&&&& 5800 & 4.25 & +0.01 & Gonzalez (\cite{G98})\\
	&&&&&&& 5892 & 4.27 & 0.00 & Fuhrmann et al (\cite{FPB97}) \\
	&&&&&&& 5811 & 4.09 & 0.00 & Gratton et al (\cite{GCC96})\\
	&&&&&&& 5882 & 4.34 & +0.01 & Edvardsson et al (\cite{EAGLNT93})\\
	&&&&&&& \bf{5860} &  \bf{4.28} &\bf{0.00} \\
	\hline
	\end{tabular}
	\end{center}
	\end{table}
\endlandscape

\landscape
	\begin{table}[!hb]
	\begin{center}
	\begin{tabular}{ c c c c c c c c c r l }
	\hline
	\hline
	Star name & TGMET &  $B-V$ & b-y& U-B & distance & Absolute & Teff & log g & [Fe/H] & source              \\
	          & score &      &    &     & in pc    & magnitude&      &       &        &(Teff, logg, [Fe/H]) \\
	\hline
	Sun & & 0.64-0.66  & 0.40-0.41 & 0.17-0.20 & 0 & 4.80-4.84 & 5777 & 4.44 & 0.00 & - \\
	\hline
	HD042618 & 2.857 & 0.632&  0.404 & 0.127 & 23.1 & 5.05 & 5775 & - & - & Kovtyukh et al (\cite{KSBG03})\\ 
	&&&&&&& 5653 & 4.58 & -0.16 & Reddy et al (\cite{red02})\\
	&&&&&&& \bf{5714} &  \bf{4.58} &\bf{--0.16} \\
	\hline
	HD071148 & 2.907& 0.625& 0.399 & 0.120 & 21.8 & 4.65 & 5850 & 4.25 & 0.00 & Mishenina et al  (\cite{Mish03})\\
	&&&&&&& 5716 & 4.34 & +0.03 & Chen et al (\cite{chen03})\\
	&&&&&&& 5703 & 4.46 & -0.08 & Reddy et al (\cite{red02})\\
	&&&&&&& \bf{5756} &  \bf{4.35} &\bf{-0.02} \\
	\hline
	HD186427 & 2.934 & 0.662&  0.416 & 0.200 &  21.4 & 4.56 & 5752 & 4.20 & +0.02 & Mishenina et al (\cite{Mish03})\\
	&&&&&&& 5765 & 4.46 & +0.09 & Santos et al (\cite{SIMRU03})\\
	&&&&&&& 5685 & 4.26 & +0.07 & Laws \& Gonzalez (\cite{LG01})\\
	&&&&&&& 5760 & 4.40 & +0.06 & Deliyannis et al (\cite{DCKB00})\\
	&&&&&&& 5700 & 4.35 & +0.06 & Gonzalez (\cite{G98})\\
	&&&&&&& 5773 & 4.42 & +0.06 & Feltzing \& Gustafsson (\cite{FG98})\\
	&&&&&&& 5766 & 4.29 & +0.05 & Fuhrmann et al (\cite{FPB97})\\
	&&&&&&& 5752 &-&-& di Benedetto, (\cite{dB98})\\
	&&&&&&& 5767 &-&-& Alonso et al (\cite{AAM96})\\
	&&&&&&& 5753 & 4.33 & +0.06 & Friel et al (\cite{friel93})\\
	&&&&&&& \bf{5753} &  \bf{4.35} &\bf{+0.06} \\
	\hline
	HD010145 & 3.003& 0.689&  0.421 & 0.212 & 36.7 & 4.87 & 5673 & 4.40 & -0.01 & Mishenina et al  (\cite{Mish03})\\
	&&&&&&& \bf{5673} &  \bf{4.40} &\bf{-0.01} \\
	\hline
	\end{tabular}
	\end{center}
	\end{table}
\endlandscape

\end{document}